\documentclass[twocolumn,twoside,slac_two]{revtex4}
\usepackage{graphicx}
\usepackage{fancyhdr}
\renewcommand{\u}{{\it u}}
\renewcommand{\d}{{\it d}}
\newcommand{\s}{{\it s}}
\renewcommand{\c}{{\it c}}
\renewcommand{\b}{{\it b}}

\newcommand{\be}{\begin{equation}}
\newcommand{\ee}{\end{equation}}
\begin{document}

\title{Precision Lattice Calculation of D and Ds decay constants}

%

\author{Eduardo Follana (for the HPQCD collaboration)}

\affiliation{University of Glasgow, Glasgow, UK}

\begin{abstract}
We present a determination of the decay constants of the $D$ and $D_s$
mesons from lattice QCD, each with a total error of about $2\%$,
approximately a factor of three better than previous calculations. We
have been able to achieve this through the use of a highly improved
discretization of QCD for charm quarks, coupled to gauge
configurations generated by the MILC collaboration that include the
full effect of sea \u, \d, and \s~quarks. We have results for a range
of \u/\d~masses down to $m_s/5$ and three values of the lattice
spacing, which allow us to perform accurate continuum and chiral
extrapolations. We fix the charm quark mass to give the experimental
value of the $\eta_c$ mass, and then a stringent test of our approach
is the fact that we obtain correct (and accurate) values for the mass
of the $D$ and $D_s$ mesons. We compare $f_D$ and $f_{D_s}$ with $f_K$
and $f_\pi$, and using experiment determine corresponding CKM elements
with good precision.

\end{abstract}

\maketitle

\thispagestyle{fancy}


\section{Introduction}

Precision calculations in lattice QCD play a crucial role in testing
our non-perturbative theoretical tools, by comparing the results of
the calculation with precisely measured quantities. In addition
accurate calculations of non-perturbative QCD quantities are very
important in the extraction of information from analysis of
experimental data, for example in the determination of the
Cabibbo-Kobayashi-Maskawa (CKM) matrix elements.

This is most clearly seen in the case of ``gold-plated'' processes,
for example the leptonic decay of $D_s$, $D_d$, $\pi$ and $K$
mesons. In this process the corresponding meson, with quark content $a
\bar b$ (or $\bar a b$) annihilates weakly into a W boson, with a
width given, up to calculated electromagnetic corrections
\cite{marciano, pdg06}, by:
\be 
\Gamma(P \rightarrow l \nu_l (\gamma)) = \frac{G_F^2
|V_{ab}|^2}{8\pi}f_{P}^2m_l^2m_{P}( 1-\frac{m_l^2}{m_{P}^2}) ^2.
\label{eq:gamma}
\ee
$V_{ab}$ is the corresponding element of the CKM matrix, and the decay
constant $f_P$ parametrizes the amplitude for W annihilation. By
combining a measurement of $\Gamma$ with an accurate calculation of
$f_P$ (\ref{eq:gamma}) can be used to determine $V_{ab}$. If $V_{ab}$
is known from elsewhere we can use (\ref{eq:gamma}) to get a value for
$f_P$.

The decay constant $f_{P}$ is conventionally defined to be a property
of the pseudoscalar meson, calculable in QCD without QED effects, and
is given by:
\begin{equation}
\langle 0 | \overline{a} \gamma_{\mu} \gamma_5 b | P(p) \rangle = f_{P} p_{\mu}.
\label{eq:fdef}
\end{equation}

The calculation of $f_P$ is a hard non-perturbative problem, which at
present can only be done fully with lattice QCD. There are very
precise experimental measurements for the leptonic decay rates in the
case of the $\pi$ and $K$, and new results are appearing for $D$ and
$D_s$, which make the calculations a highly non-trivial test of
lattice QCD, and ultimately of QCD itself. This tests are important to
give us confidence in similar lattice QCD predictions of matrix
elements in B systems, for which experimental results are much harder
to obtain.

\section{Improved Staggered Quarks}

We use HISQ staggered quarks in the valence sector, whereas the sea
quarks are ASQTAD staggered quarks with the fourth root trick
\cite{Sharpe, Creutz, Kronfeld}.

Staggered quark actions suffer the doubling problem: there are four
``tastes'' (non-physical flavours) of fermions in the spectrum, which
couple through taste-changing interactions. These are lattice
artifacts of order $a^2$, involving at leading order the exchange of a
gluon of momentum $q \approx \pi/a$.  Although quite large in the
original one-link (Kogut-Susskind) staggered action, such interactions
are perturbative for typical values of the lattice spacing, and can be
corrected systematically a la Symanzik. By judiciously smearing the
gauge field we can remove the coupling between quarks and high
momentum gluons.

The most widely used improved staggered action is called ASQTAD, and
removes all tree-level $a^2$ discretization errors in the action
\cite{Naik, Lepage1, Orginos}. 

The HISQ (highly improved staggered quarks) staggered Dirac operator
involves two levels of smearing with an intermediate projection onto
$SU(3)$. It is designed so that, as well as eliminating all tree-level
$a^2$ discretization errors, it further reduces the one-loop
taste-changing errors (see \cite{HISQ} for a more detailed
discussion.) This action has been shown to substantially reduce the
errors associated with the taste-changing interactions \cite{HISQ,
spectrum1, spectrum2}.

When we put massive quarks on the lattice, the discretization errors
grow with the quark mass as powers of $a m$. Therefore to obtain small
errors we would need $a m \ll 1$. For heavy quarks this would require
very small lattice spacings. On the other hand, to keep our lattice
big enough to accommodate the light degrees of freedom, we need $L a
\gg m_\pi^{-1}$. The fact that we have two very different scales in
the problem makes difficult a direct solution. What we can do instead
is to take advantage of the fact that $m$ is large, by using an
effective field theory (NRQCD, HQET). This program has been very
successful for \b~quarks \cite{NRQCD1, NRQCD2, FNAL}.

The charm quark is in between the light and heavy mass regime. It is
quite light for an easy application of NRQCD, but quite large for the
usual relativistic quark actions, $a m_c \stackrel {\textstyle
<}{\sim} 1$. However, if we use a very accurate action (HISQ) and fine
enough lattices (fine MILC ensembles), it is possible to get results
accurate at the few percent level. A non-relativistic analysis
\cite{HISQ} shows that for HISQ charm quarks the largest remaining
source of error is due to the quark's energy, and can be further
suppressed by powers of $v/c$, where $v$ is the typical velocity of
the quark in the system of interest, simply by retuning the overall
coefficient of a term called Naik term to impose the correct
relativistic dispersion relation $c^2(p) = 1$ for low lattice momemtum
$p$.

One advantage of the use of a relativistic action is the existence of
a partially conserved current, which implies the non-renormalization
of the lattice result for $f_P$. We can extract $f_P$ from the PCAC
relation for zero momentum meson P:
\be
f_P m_P^2 = (m_a + m_b) \left<0|\bar a \gamma_5 b|P\right>
\ee

\section{Results}

\begin{table}[h]
\begin{tabular}{l|l|l}
Lattice/sea & valence & $r_1/a$ \\
$u_0am_l$, $u_0am_s$ & $am_l$, \,\,\, $am_s$, \,\, $am_c$, $1+\epsilon$ & \\
\hline
$16^3 \times 48$ & & \\
0.0194, 0.0484 & 0.0264, 0.066, 0.85, 0.66 & 2.129(11) \\
0.0097, 0.0484 & 0.0132, 0.066, 0.85, 0.66 & 2.133(11) \\
$20^3 \times 64$ & & \\
0.02, 0.05 & 0.0278, 0.0525, 0.648, 0.79 & 2.650(8) \\
0.01, 0.05 & 0.01365, 0.0546, 0.66, 0.79 & 2.610(12) \\
$24^3 \times 64$ & & \\
0.005, 0.05 & 0.0067, 0.0537, 0.65, 0.79 &  2.632(13) \\
$28^3 \times 96$ & & \\
0.0124, 0.031 & 0.01635, 0.03635, 0.427, 0.885 & 3.711(13) \\
0.0062, 0.031 & 0.00705, 0.0366, 0.43, 0.885 & 3.684(12) \\
\end{tabular}
\caption{MILC configurations and mass parameters used for this
analysis. The $16^3 \times 48$ lattices are `very coarse', the $20^3
\times 64$ and the $24^3 \times 64$, `coarse' and the $28^3 \times
96$, `fine'. The sea asqtad quark masses ($l = u/d$) are given in the
MILC convention where $u_0$ is the plaquette tadpole parameter. Note
that the sea \s~quark masses on fine and coarse lattices are above the
subsequently determined physical value~\cite{MILC3}.  The lattice
spacing values in units of $r_1$ after `smoothing' are given in the
rightmost column~\cite{MILC2, priv}. The third column gives the HISQ
valence \u/\d, \s~and \c~masses along with the coefficient of the Naik
term, $1+\epsilon$, used for \c~quarks~\cite{HISQ}. }
\label{tab:params}
\end{table}

We use gluon field configurations including $2+1$ flavours of sea
quarks generated by the MILC collaboration \cite{MILC1, MILC2,
MILC3}. The parameters of the ensembles we have used for both the sea
and the valence sectors are in table \ref{tab:params}. The lattice
results are converted to physical units through the heavy quark
potential parameter $r_1$, as determined by the MILC collaboration
(table \ref{tab:params}, \cite{MILC2}). The physical value of $r_1$ is
determined from the $\Upsilon$ spectrum calculated in NRQCD with
\b~quarks on the same MILC ensembles \cite{NRQCD2}, with the result
$r_1 = 0.321(5)$ fm, $r_1^{-1} = 0.615(10)$ GeV.

We use multiple precessing random wall sources, which gives a 3-4-fold
reduction in statistical errors with respect to conventional local
sources.

\begin{figure}
\center{
  \includegraphics[width=.65\hsize,angle=-90]{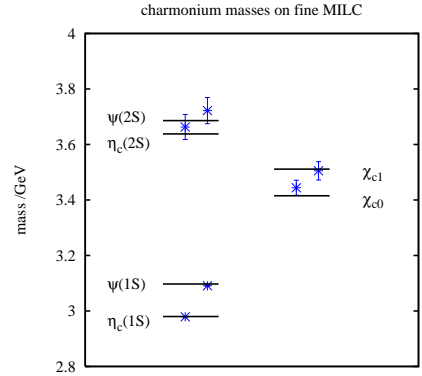}
}
\caption{\label{hisqspect} Charmonium spectrum obtained with HISQ
charm quarks on fine MILC lattices (blue crosses with error bars)
against experiment (black lines).}
\end{figure}

The mass of the charm quark is fixed by adjusting the mass of the
``goldstone `` $\eta_c$ to its experimental value. The light (\u/\d)
and strange quark masses are fixed using the experimental values for
the masses of $\pi$ and $K$. Our results use masses for the \u~and
\d~quarks that are substantially larger (by a factor of around three)
than the real ones. In order to get physical answers we extrapolate to
the correct \u/\d~mass using chiral perturbation theory. Once the
masses have been thus fixed, there is no remaining freedom to change
any parameters, and in particular the results we obtain for the masses
of heavy-light mesons are a stringent test of our method. In figure
\ref{hisqspect} we show the spectrum of charmonium. We obtain an
hyperfine splitting of $111(5)$ MeV (experiment 117(1) MeV) We have
made no attemp as yet to optimize the calculation of the excited
states.

\begin{figure}
\center{
  \includegraphics[width=.8\hsize]{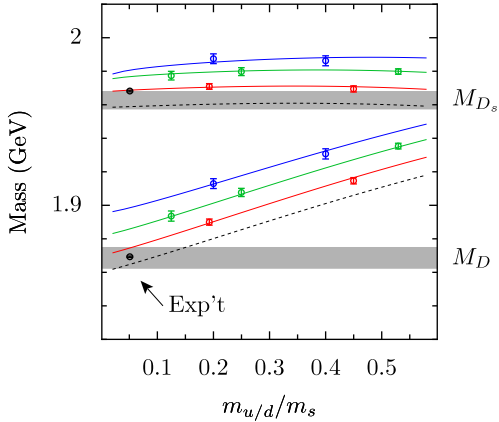}
}
\caption{\label{md} Masses of the $D^+$ and $D_s$ meson as a function
of the \u/\d~quark mass in units of the \s~quark mass at
three different values of the lattice spacing. The very coarse results
are above the coarse and the fine are the lowest. The lines give the
simultaneous chiral fits and the dashed line the continuum
extrapolation as described in the text. Our final error bars,
including the overall scale uncertainty, are given by the shaded
bands. These are offset from the dashed lines by an estimate of
electromagnetic, $m_{u}\ne m_{d}$ and other systematic corrections to
the masses. The experimental results are marked at the physical
$m_{d}/m_s$.  }
\end{figure}

\begin{figure}
\center{
  \includegraphics[width=.8\hsize]{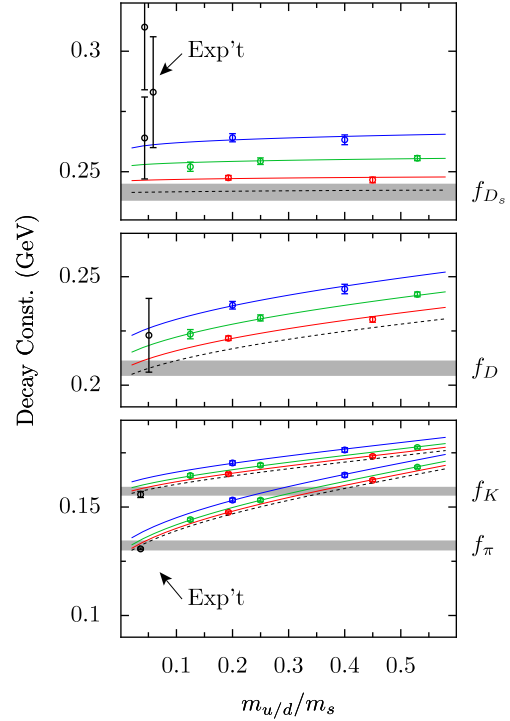}
}
\caption{ Results for the $D$, $D_s$, $K$ and $\pi$ decay constants on
very coarse, coarse and fine ensembles, as a function of the $u/d$
quark mass in units of the \s~quark mass.  The chiral fits are
performed simultaneously with those of the corresponding meson masses,
and the resulting continuum extrapolation curve is given by the dashed
line. The shaded band gives our final result.  At the left are
experimental results from CLEO-c~\cite{cleocfds, cleocfd} (on the left
with the $\tau$ decay result above the $\mu$ decay result for $D_s$)
and BaBar~\cite{babar} ($D_s$ only) and from the Particle Data
Tables~\cite{pdg06} for $K$ and $\pi$. For the $K$ we have updated the
result quoted by the PDG to be consistent with their quoted value of
$V_{us}$.}
\label{fd}
\end{figure}

In addition to the chiral extrapolation, we have systematic errors
coming from a variety of sources \cite{fds}, among them from the
finite lattice spacing. Because we have three different lattice
spacings and very precise data, we can extrapolate to the continuum
limit. This extrapolation is linked to the chiral extrapolation
through discretization errors in the light quark action. We therefore
perform a simultaneous bayesian fit for both chiral and continuum
extrapolations, allowing for expected functional forms in both. We
tested the validity of the method by fitting hundreds of fake datasets
generated using staggered chiral perturbation theory with random
couplings. We fit simultaneously to the masses and the decay
constants, that is, we fit $m_\pi$, $m_K$, $f_\pi$ and $f_K$
simultaneously, and similarly for $m_D$, $m_{D_s}$, $f_D$ and
$f_{D_s}$. We present some of the results in figures \ref{md} and
\ref{fd}.
\begin{figure}
\includegraphics[width=.8\hsize,angle=-90]{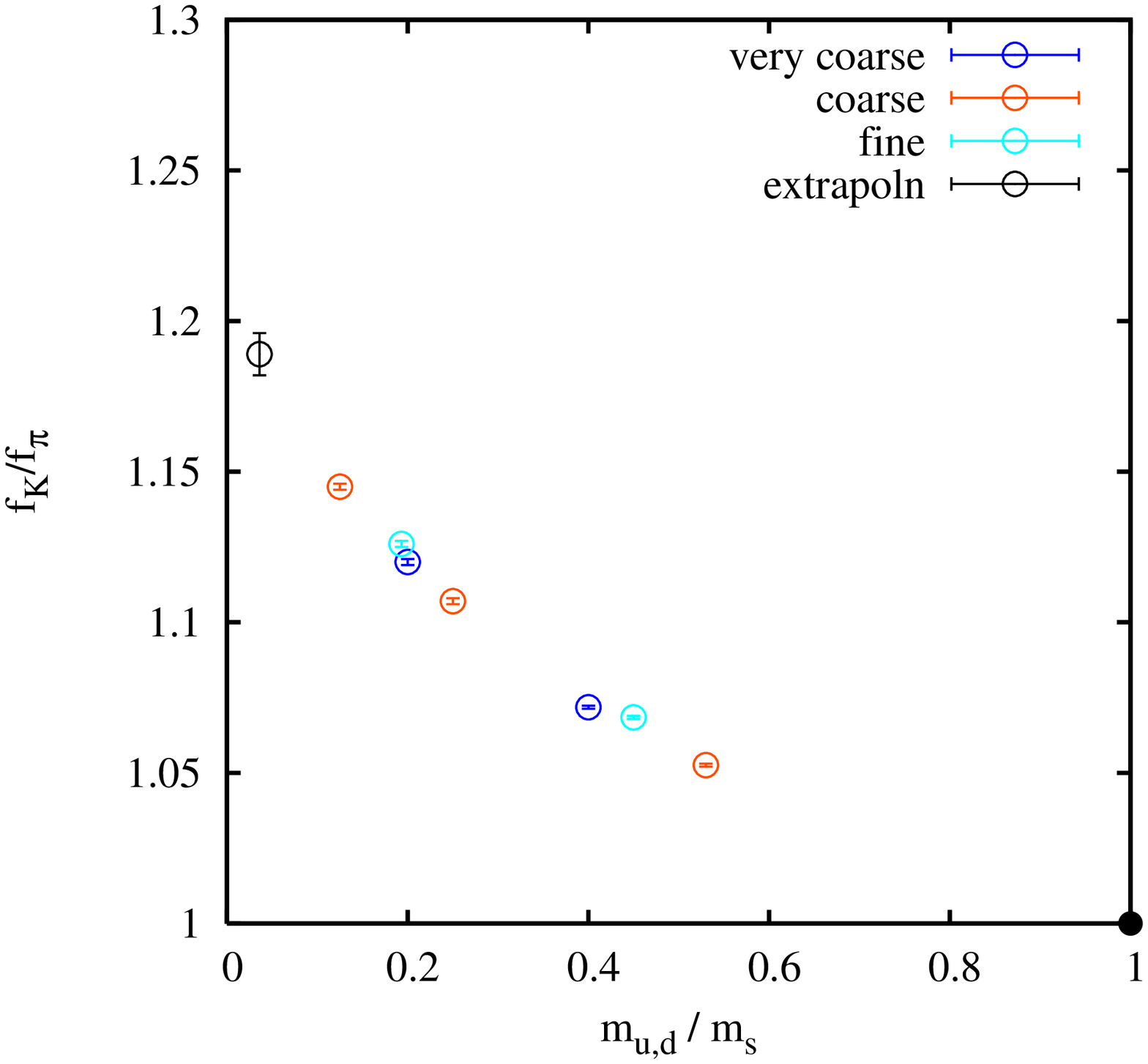}
\caption{Ratio of decay constants $f_K / f_\pi$ on very coarse, coarse
and fine ensembles, as a function of the \u, \d~quark mass in units
of the \s~quark mass.}
\label{fpi_fk}
\end{figure}
\begin{figure}
\includegraphics[width=.8\hsize,angle=-90]{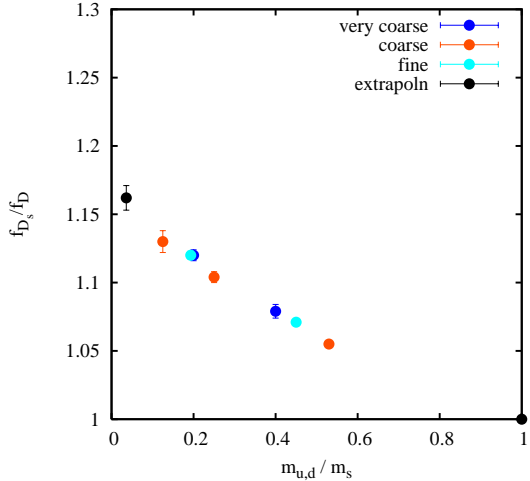}
\caption{Ratio of heavy-light decay constants $f_{D_s}/f_D$ on very
coarse, coarse and fine ensembles, as a function of the \u, \d~quark
mass in units of the \s~quark mass.}
\label{fd_fds}
\end{figure}

We get an excellent agreement with experiment for the masses: $m_{D_s}
= 1.963(5)$ GeV (experiment $1.968$ GeV), and $m_D = 1.869(6)$ GeV
(experiment $1.869$ GeV). Our calculation also reproduces correctly
the difference in binding energies between a heavy-heavy ($\eta_c$)
and a heavy-light ($m_D$ and $m_{D_s}$) state: $(2 m_{D_s} -
m_{\eta_c}) / (2 m_D - m_{\eta_c}) = 1.249(14)$ (experiment
$1.260(2)$). Our charm quark action is the first one to be accurate
enough to do this calculation (which also cannot be done, for example,
in potential models.)

We also have agreement with experiment for the light-light decay
constants \cite{fds}. The result for the ratio is very accurate,
$f_K/f_\pi = 1.189(7)$, and shows tiny discretization effects (figure
\ref{fpi_fk}). Combining this ratio with experimental leptonic
branching fractions \cite{MILC3,marciano2} we get $V_{us} =
0.2262(13)(4)$, where the first error is theoretical and the second
experimental. This gives the unitarity relation
$1-V_{ud}^2-V_{us}^2-V_{ub}^2 = 0.0006(8)$.

Our results for the heavy-light decay constants are 4-5 times more
accurate than previous lattice QCD results and existing experimental
measurements: $f_{D_s} = 241(3)$ MeV, $f_D = 208(4)$ MeV, and a ratio
of $f_{D_s}/f_D = 1.162(9)$ (see figure \ref{fd_fds}). For the double
ratio $(f_{D_s}/f_D)/(f_K/f_{\pi})$, which is estimated to be close to
1 from low order chiral perturbation theory~\cite{becirevic}, we get a
value of $0.977(10)$.

The experimental leptonic branching rates, together with CKM matrix
elements determined from other processes (assuming $V_{cs} = V_{ud}$)
give $f_{D_s} = 264(17)$ MeV for $\mu$ decays and $310(26)$ MeV for
$\tau$ decay from CLEO-c~\cite{cleocfds} and 283(23) MeV from
BaBar~\cite{babar}, and for $f_D$ 223(17) MeV from CLEO-c for $\mu$
decay~\cite{cleocfd}. Using our results for $f_{D_s}$ and
$f_{D_s}/f_D$ and the experimental values from CLEO-c~\cite{cleocfds}
for $\mu$ decay (since the electromagnetic corrections are well-known
in that case) we can directly determine the corresponding CKM
elements: $V_{cs} = 1.07(1)(7)$ and $V_{cs}/V_{cd}$ = 4.42(4)(41).
The first error is theoretical and the second experimental.  The
result for $V_{cs}$ improves on the direct determination of 0.96(9)
given in the Particle Data Tables~\cite{pdg06}.

\begin{figure}
\includegraphics[width=.85\hsize,angle=-90]{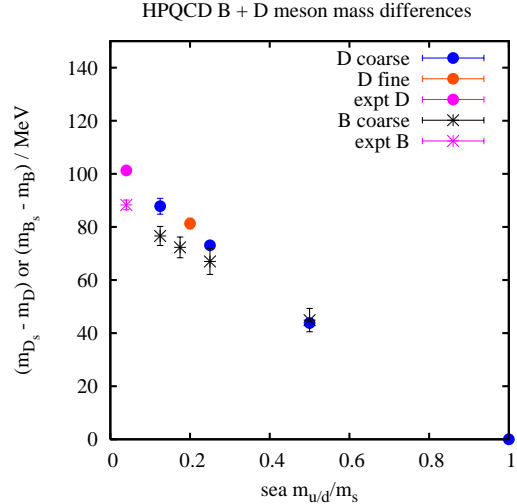}
\caption{Binding energy differences of heavy-light mesons with strange
valence quarks and those with \u, \d~valence quarks, for states with a
\c~quark and systems with a \b~quark, $m_{D_s} - m_D$ and $m_{B_s} -
m_B$. Data are shown for the coarse and fine ensembles in the case of
$D$ and $D_s$, and for the coarse for the $B$ and $B_s$.  }
\label{mdiff}
\end{figure}

Our calculation is precise enough that we can see the difference
between $m_{B_s}(m_l) - m_B(m_l)$ in the bottom sector and the similar
quantity in the charm sector $m_{D_s}(m_l) - m_D(m_l)$ (figure
\ref{mdiff}). These mass differences are small compared to the
absolute masses of the states, and should be the same in the
infinitely heavy quark limit. We can see that our calculation
correctly reproduces the small difference due to the finite value of
the mass of the charm and bottom.

\section{Conclusions and outlook}

We have shown that the use of a highly improved relativistic action on
fine enough lattices is capable of delivering very precise results on
systems with a charm quark. The high statistical accuracy of our data
combined with calculations at several values of the lattice spacing
and light quark masses allows us to make a controlled joint chiral and
continuum extrapolation.

We can calculate accurately the mass of heavy-light systems, which
provide a stringent test of the calculation. We can calculate precise
values for the decay constants of pseudoscalar heavy-light mesons (as
well as light-light mesons), and especially for the ratio of such
decay constants.

The very precise calculation of the masses of heavy-heavy pseudoscalar
mesons should make possible a direct lattice determination of the mass
of the charm quark. Because we use the same relativistic action
through the calculation for both the charm and the light quarks, we
can also obtain a very precise value for the ratio $m_c / m_s$, and
therefore if $m_c$ is determined through another method use the ratio
to get $m_s$. We are also working (in collaboration with the Karlsruhe
group) on a new method for the determination of $m_c$ by combining
continuum perturbation results with lattice data.

Another quantity which we plan to calculate in the near future is the
leptonic decay width $\Gamma_{e^+e^-} (\psi)$, as well as the
semileptonic form factors for $D \rightarrow \pi l \nu$, $D
\rightarrow K l \nu$.

\begin{acknowledgments}
We are grateful to the MILC collaboration for the use of their
configurations and to Quentin Mason and Doug Toussaint for useful
discussions. The computing was done on Scotgrid and the QCDOCX
cluster.  This work was supported by PPARC, the Royal Society, NSF and
DoE.
\end{acknowledgments}

\bigskip 

\end{document}